%% file: audio_ssl.tex
\documentclass[conference]{IEEEtran}
\IEEEoverridecommandlockouts
\usepackage{cite}
\usepackage{amsmath,amssymb,amsfonts}
\usepackage{algorithmic}
\usepackage{graphicx}
\usepackage{textcomp}
\usepackage{tabularx}
\usepackage{xcolor}
\usepackage{paralist}
\usepackage{units}
\usepackage{bm}
\usepackage{microtype}

\newcolumntype{b}{X}
\newcolumntype{s}{>{\hsize=.5\hsize}X}
\newcommand{\heading}[1]{\multicolumn{1}{c}{#1}}

\input{math.tex}

\def\BibTeX{{\rm B\kern-.05em{\sc i\kern-.025em b}\kern-.08em
    T\kern-.1667em\lower.7ex\hbox{E}\kern-.125emX}}
\begin{document}

\title{Semi-Supervised Audio Classification with Partially Labeled Data}

\author{\IEEEauthorblockN{Siddharth Gururani}
\IEEEauthorblockA{
\textit{Georgia Institute of Technology}\\
siddgururani@gatech.edu}
\and
\IEEEauthorblockN{Alexander Lerch}
\IEEEauthorblockA{
\textit{Georgia Institute of Technology}\\
alexander.lerch@gatech.edu}
}

\maketitle

\begin{abstract}
    Audio classification has seen great progress with the increasing availability of large-scale datasets. These large datasets, however, are often only partially labeled as collecting full annotations is a tedious and expensive process. This paper presents two semi-supervised methods capable of learning with missing labels and evaluates them on two publicly available, partially labeled datasets. The first method relies on label enhancement by a two-stage teacher-student learning process, while the second method utilizes the mean teacher semi-supervised learning algorithm. Our results demonstrate the impact of improperly handling missing labels and compare the benefits of using different strategies leveraging data with few labels. Methods capable of learning with partially labeled data have the potential to improve models for audio classification by utilizing even larger amounts of data without the need for complete annotations.
\end{abstract}


\section{Introduction}
\label{sec:intro}


Automatic audio classification is at the core of a wide variety of tasks in audio analysis and music information retrieval (MIR). 
Improvements in audio classification models, especially deep neural network-based models, triggered the creation of large-scale datasets for audio classification, both for general purpose audio~\cite{gemmekeAudio2017} as well as for specific domains, such as musical instruments~\cite{humphreyOpenMIC20182018}. However, the larger the scale of data, the more difficult it is to annotate. One way for annotating large-scale datasets is by using services such as Mechanical Turk~\cite{humphreyOpenMIC20182018}. However, for crowd-sourced annotations, it may be impractical to annotate both the presence (positives) and absence (negatives) of each class for a large number of classes and audio clips. Thus, audio clips are often annotated partially, i.e., each clip may only be annotated with presence/absence of a few classes out of the set of classes.

\begin{figure}[t]
  \begin{minipage}[b]{1.0\linewidth}
    \centering
    \centerline{\includegraphics[width=.7\columnwidth]{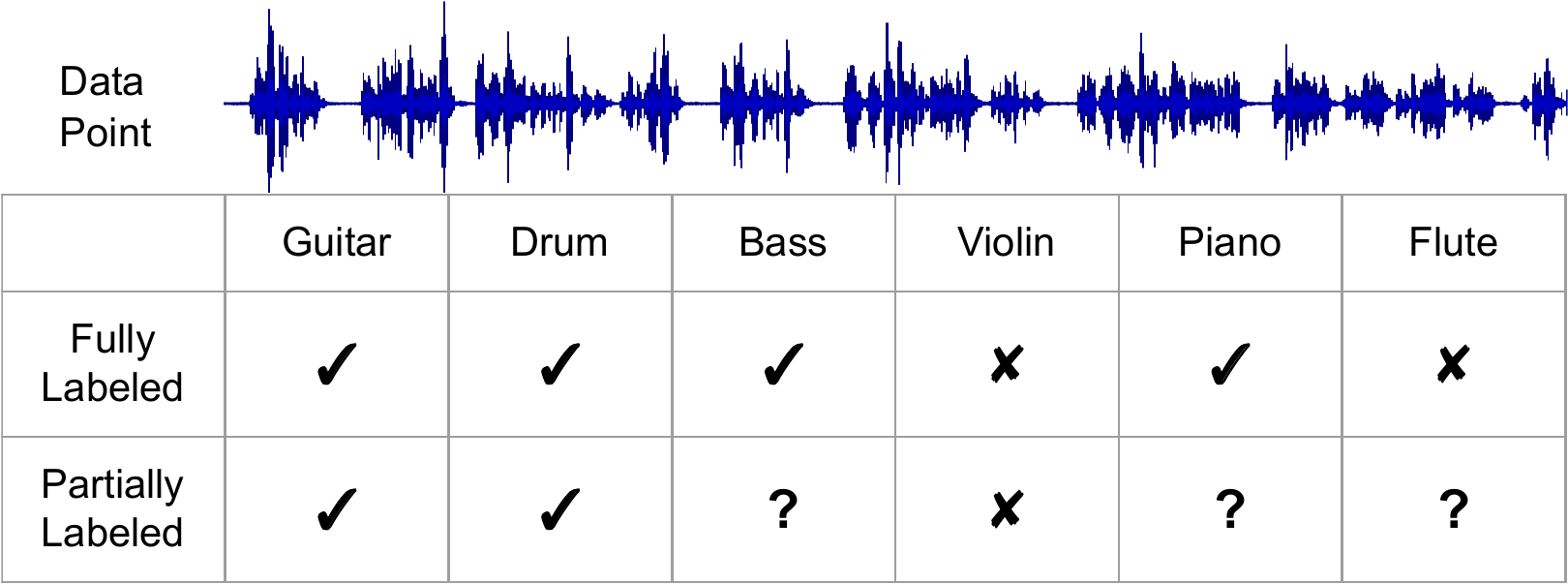}}
    \caption{An example of a music clip along with complete and partial labels.}
    \label{label_types}
  \end{minipage}
\end{figure}

We focus on the problem of missing labels in audio classification datasets. \Figref{label_types} illustrates the difference between partially labeled and fully labeled data. The conventional way to handle data with incomplete annotations is to treat missing labels as negative instances for the respective class (compare, e.g., \cite{fonsecaAddressing2020a,hersheyCNN2017}) since there is no obvious alternative way to handle this. The Audioset~\cite{gemmekeAudio2017}, for example, a large audio dataset available for audio classification, labels each clip only with a few `present' class labels. It is, however, unclear whether all of the remaining classes are indeed `absent.' As pointed out by Fonseca et al., this is problematic since the missing labels can mean either present or absent instances~\cite{fonsecaAddressing2020a}. 

For example, an audio clip which contains a `trumpet' and a `saxophone' may only be labeled with the trumpet class. Treating the missing saxophone label as negative (no `saxophone') represents a mislabeled example from the point-of-view of an audio classifier and may negatively impact its training. 
Thus, it is important to investigate approaches capable of handling data with missing labels. Moreover, methods capable of leveraging missing labels enable composition of larger datasets by combining existing datasets even if their label sets may not fully overlap.
The main contributions of the paper are as follows:
\begin{inparaenum}
  \item We identify a problem in music and audio tagging tasks which is often overlooked in research tackling the respective task: existing methods do not account appropriately for partially labeled data.
  \item We compare two semi-supervised learning approaches to handle missing labels. 
  In contrast to other works on semi-supervised audio classification, which divide datasets into labeled and unlabeled, we deal with homogenous multi-label datasets where each data point has at least one class label.
  \item We evaluate these methods using large publicly available datasets from different domains: OpenMIC, and SONYC-UST. 
These datasets are both partially labeled and consist of music and urban sound recordings, respectively, thus helping us understand the behavior and effectiveness of the presented methods on diverse sets of audio.
\end{inparaenum}


\section{Related Work}
\label{sec:format}


Training models with missing labels has been the subject of several machine learning studies. Semi-supervised learning (SSL) is particularly useful where there is limited labeled data. 
Most SSL work tends to utilize large image datasets such as CIFAR100, or SVHN~\cite{tarvainenMean2018}. 
Typically, labeled and unlabeled partitions are created ad hoc by discarding labels from images in the unlabeled set.
Recent work in sound event detection has also applied SSL approaches in a similar fashion, with the goal to improve models trained with small fully labeled data and large unlabeled data~\cite{akiyamaMultitask2019,luSemiSupervised2019}. Similarly, Kum et al.\ presented variations of noisy student-teacher learning approaches for vocal melody extraction~\cite{keum2020semi}. They utilized a small labeled dataset combined with a larger unlabeled dataset for this task.
Our paper, on the other hand, focuses on data from the same collection or dataset where each sample is partially labeled. This setting represents the state of various large-scale audio tagging datasets and is hence an important problem to solve. 

Only a few studies address the problem of missing labels in audio classification.
Meire et al.\ studied the effect of varying degrees of missing labels for matrix factorization and convolutional neural network (CNN)-based models~\cite{meireImpact2019}. They found that the CNN failed to outperform the NMF-based approach for cases with few labeled samples in a synthetic dataset.
Fonseca et al.\ utilized the student-teacher paradigm and a masked binary cross-entropy (BCE) loss to address missing labels using a proprietary version of the Audioset~\cite{fonsecaAddressing2020a} with positive and explicit negative labels.
Gururani et al.\ utilized a partial BCE loss to ignore missing labels in the OpenMIC dataset while training models for musical instrument classification~\cite{gururaniAttention2019}. While they address missing labels, they only do so by omission.

\section{Partially Labeled Datasets}
\label{sec:pagestyle}
The main reason for missing labels in audio datasets is the difficulty of annotating audio. It is non-trivial for the human ear to instantly identify several sound events in a single audio clip. This is in contrast to labeling an image where salient objects can be easily identified with a quick scan. The problem is exacerbated for music recordings since most instruments overlap in time and may have similarities in timbre.

A typical setup for audio annotation involves asking annotators whether a certain class of sound event or musical instrument is active in a given audio clip~\cite{fonsecaFSD50K2020, humphreyOpenMIC20182018}. A listener has to listen to the entire audio clip in order to annotate. Depending on the length of the clip, this may take a few seconds to several minutes. This drastically increases the cost of annotations. This challenge can be overcome for specific data (e.g., letting the annotators skip silent parts in the case of urban or environmental sounds), but is not possible in the case of music where audio clips typically have active instruments for a majority of the time. These constraints lead to only a limited amount of annotations for each audio clip in large-scale multi-label audio datasets.

We choose two partially labeled datasets for our experiments. Both datasets share the property that they have positive and negative labels for each class. The negative labels are equivalent to `explicit negatives' as referred to in \cite{fonsecaAddressing2020a}. 



The SONYC Urban Sound Tagging (SONYC-UST) dataset\footnote{Version 2.3.0, DOI 10.5281} is a dataset of urban sounds collected using an acoustic sensor network in New York City \cite{cartwrightSONYC2020}. The dataset consists of $13538$ training, $4308$ validation, and $669$ test recordings, each of \unit[10]{s} length. Each recording is annotated with the presence or absence of $23$ classes of urban noise. The SONYC-UST is approximately $94\%$ labeled. 


The OpenMIC dataset is a large-scale, multi-instrument music dataset with a diverse set of instruments and genres, addressing issues in other datasets for the task~\cite{humphreyOpenMIC20182018}. It consists of $20000$ \unit[10]{s} audio clips. Each clip is extracted from a unique song in the Free Music Archive~\cite{defferrard2017fma}. Each clip is labeled with the presence or absence of least one of $20$ musical instruments. The dataset is weakly labeled implying that instrument labels do not indicate the precise activity in time of the instrument in the audio clip. If the dataset was completely labeled, the number of labels would be $20000 \text{ clips} \cdot 20 \text{ instruments} = 400000$. However, the number of labels in the dataset is $41268$, showing that approximately $90\%$ of the labels are missing.


\section{Methods} \label{sec:method}

We present two methods geared towards addressing the challenge of missing labels. These methods belong to different categories of semi-supervised learning algorithms. The first method can be loosely categorized as self-training while the second method is based on consistency regularization.

\subsection{Method 1: Label Enhancing (LE)} \label{sec:method_LE}
\begin{figure}[!t]
  \begin{minipage}[b]{1.0\linewidth}
    \centering
    \centerline{\includegraphics[width=6.5cm]{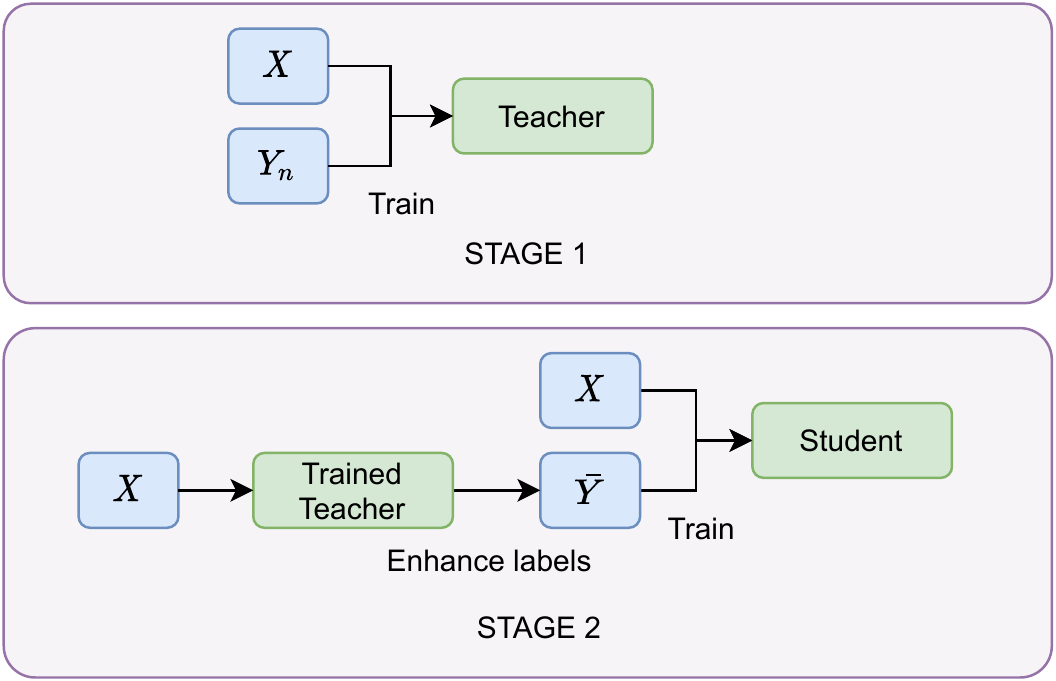}}
    \caption{Overview of the two stage Label Enhancing (LE) method.}
    \label{LE_arch}
  \end{minipage}
\end{figure}

Label Enhancing is a method proposed by Fonseca et al.~\cite{fonsecaAddressing2020a}. LE involves two stages of model training. In the first stage, a teacher model is trained using the label set $Y_n$ such that all missing labels (implicit negatives) are treated as explicit negative labels. In the second stage, an enhanced label set $\bar{Y}$ is obtained using the teacher model's predictions $S$ for the missing labels. \Figref{LE_arch} illustrates the LE algorithm. In contrast to traditional self-training where missing labels are replaced by model predictions, here the enhanced label set is created by masking out certain implicit negatives during training of the student model. The mask is generated as follows:
\begin{align}
    \mathcal{M}_c = \begin{cases}
        0, & \text{if label is missing and $S_c >=\tau$ }\\
        1, & \text{otherwise},
    \end{cases}
\end{align}
where $\mathcal{M}_c$ and $S_c$ denote class-conditional mask and model predictions, respectively. The threshold $\tau$ is obtained using $\gamma$ percentile of teacher predictions for missing labels in each class. The authors argue that the implicit negatives with a high value for model prediction are likely to be missing positives. Thus, removal of these labels forms an enhanced label set $\bar{Y}$, which is finally used to train a student model. 

\subsection{Method 2: Mean Teacher (MT)}

MT is a popular consistency regularization-based (CR) semi-supervised learning (SSL) algorithm. CR-based SSL works on the principle that a model should output consistent predictions for perturbed or noisy versions of the same input. These perturbations can vary from simple dropout~\cite{bachmanLearning2014} to various data augmentation methods~\cite{berthelotMixMatch2019,sajjadiRegularization2016a}. 


Formally, the consistency loss term can be written as~\cite{tarvainenMean2018}:
$
    \mathcal{J}_\mathrm{CR} = \expect{\rvx, \eta, \eta'}{\mathcal{D}(f(\mathcal{T}(x), \theta, \eta), f(\mathcal{T}(x), \theta', \eta'))}, \label{J_CR}
$
where $\mathcal{D}$ is some distance function, $f$ represents the computational graph of a deep neural network with parameters $\theta$ and stochastic noise (dropout) $\eta$, and $\mathcal{T}$ is some data transformation or augmentation function to inject noise in training data. The network with parameters $\theta$ is the student model, and the network with parameters $\theta'$ is the teacher model. In our experiments, we only utilize dropout as the source of stochastic noise.
In MT, the teacher model weights $\theta$ are the exponential moving average (EMA) of the student model weights $\theta$ after every training step: 
$\theta'_t = \alpha\theta'_{t-1} + (1-\alpha)\theta_{t}$
where $\alpha$ is the EMA weight.

\begin{figure}[!t]
  \begin{minipage}[b]{1.0\linewidth}
    \centering
    \centerline{\includegraphics[width=.6\columnwidth]{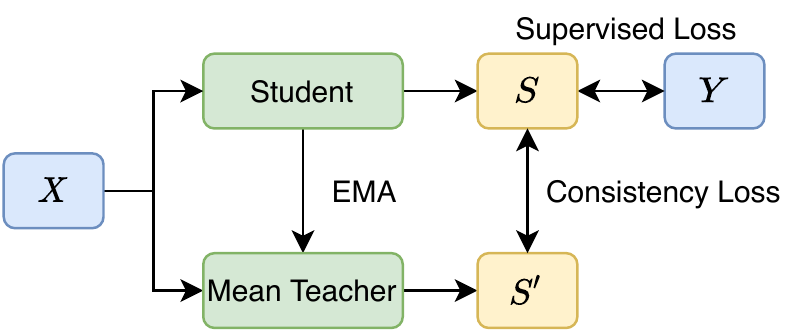}}
    \caption{Overview of Mean Teacher (MT) approach.}
    \label{MT_arch}
  \end{minipage}
\end{figure}

Thus, the student model is trained with both:
\begin{inparaenum}[(i)]
    \item the BCE loss $\mathcal{J}_\mathrm{BCE}$ which utilizes the labeled data, and
    \item the consistency loss which does not require labels and is well suited to data with missing labels.
\end{inparaenum} The final loss function is:
  $\mathcal{J} = \mathcal{J}_\mathrm{BCE} + \beta\ \mathcal{J}_\mathrm{CR}$,
where $\beta$ is a hyperparameter that decides the weight of the consistency regularization term.
Meanwhile, the teacher aggregates the student model by averaging the student weights as the training progresses. \Figref{MT_arch} illustrates the MT method.

\begin{table}[t]
  \centering
    \caption{L: number of hidden layers, H: hidden size, D: dropout}
    \label{tab:hparams}
    \begin{tabularx}{\columnwidth}{s|b|b}
      \heading{ }  & \heading{OpenMIC}                            & \heading{SONYC-UST}                          \\ \hline
    Model   & L: 3, H: 128, D: 0.6               & L: 1, H: 512, D: 0.6                \\ \hline
    Optimizer     & \multicolumn{2}{c}{lr: \{0.001, $5e^{-3}$\}, weight decay: $1e^{-5}$}          \\ \hline
    Other & $\gamma$: 10\%, $\alpha$: 0.999, $\beta$: 3 & $\gamma$: 95\%, $\alpha$: 0.99, $\beta$: 3
    \end{tabularx}
\end{table}

\section{Experimental Setup}
\label{sec:exp}

We train our models to identify presence or absence of sound classes: musical instruments in case of OpenMIC, and urban sounds in case of SONYC-UST. These are multi-label classification problems. The setup for instrument classification closely follows the study by Gururani et al.~\cite{gururaniAttention2019}. We use the macro-average F1-score to evaluate performance which takes into account data imbalances.
The setup for urban sound tagging is the same as the baseline system for DCASE 2019 Challenge Task 5.\footnote{http://dcase.community/challenge2019/task-urban-sound-tagging} 
For evaluation, we use the micro- and macro-average AUPRC to be consistent with the challenge.
Code to reproduce the experiments is made publicly available.\footnote{https://github.com/SiddGururani/audio\_missing\_labels}

\subsection{Baseline Systems}\label{baselines}

\subsubsection{Baseline 0 (B0)}
Missing labels are treated as negative labels. This method is common in audio classification research involving datasets like the AudioSet~\cite{kongAudio2018} since it only contains positive/presence labels~\cite{fonsecaAddressing2020a}. It is often a forced choice due to the data collection procedure. Additionally, since the focus is usually on devising improved acoustic models, the nuances of the label collection process tend to be ignored.

\subsubsection{Baseline 1 (B1)}
Missing labels are ignored or masked in the loss function during training, a naive way of handling missing labels~\cite{gururaniAttention2019}. Compared to B0, this protects against learning from incorrect negative labels. Loss masking is utilized in the LE method during the training of the student model, as well as in the MT method for the supervised loss function. 

\subsection{Model Training Pipeline}

Pre-trained models extract the features used as model inputs.
The OpenMIC data is processed using the VGGish model to obtain 128-dimensional embeddings for every \unit[0.96]{s} of audio~\cite{hersheyCNN2017}. Similarly, the OpenL3 library~\cite{cramerLook2019} is used to extract features for SONYC-UST. 512-dimensional embeddings are extracted for every \unit[1]{s} of audio. No geo-spatial and temporal information from SONYC-UST is utilized.
All methods presented share the same attention-based model architecture. This model solves a multiple-instance learning problem and uses attention mechanism as an aggregation/pooling strategy. It achieved state-of-the-art performance for Audioset classification as well as multi-instrument classification~\cite{kongAudio2018,gururaniAttention2019}. \Tabref{tab:hparams} displays the hyperparameters. Adam optimizer is used.
Publicly available data splits are used for training and testing. For OpenMIC, 15\% of the training data is sampled randomly to form the validation set. Validation loss is used to for model selection. 




\section{Results and Discussion}

\begin{figure}[!t]
    \centerline{\includegraphics[width=.8\columnwidth]{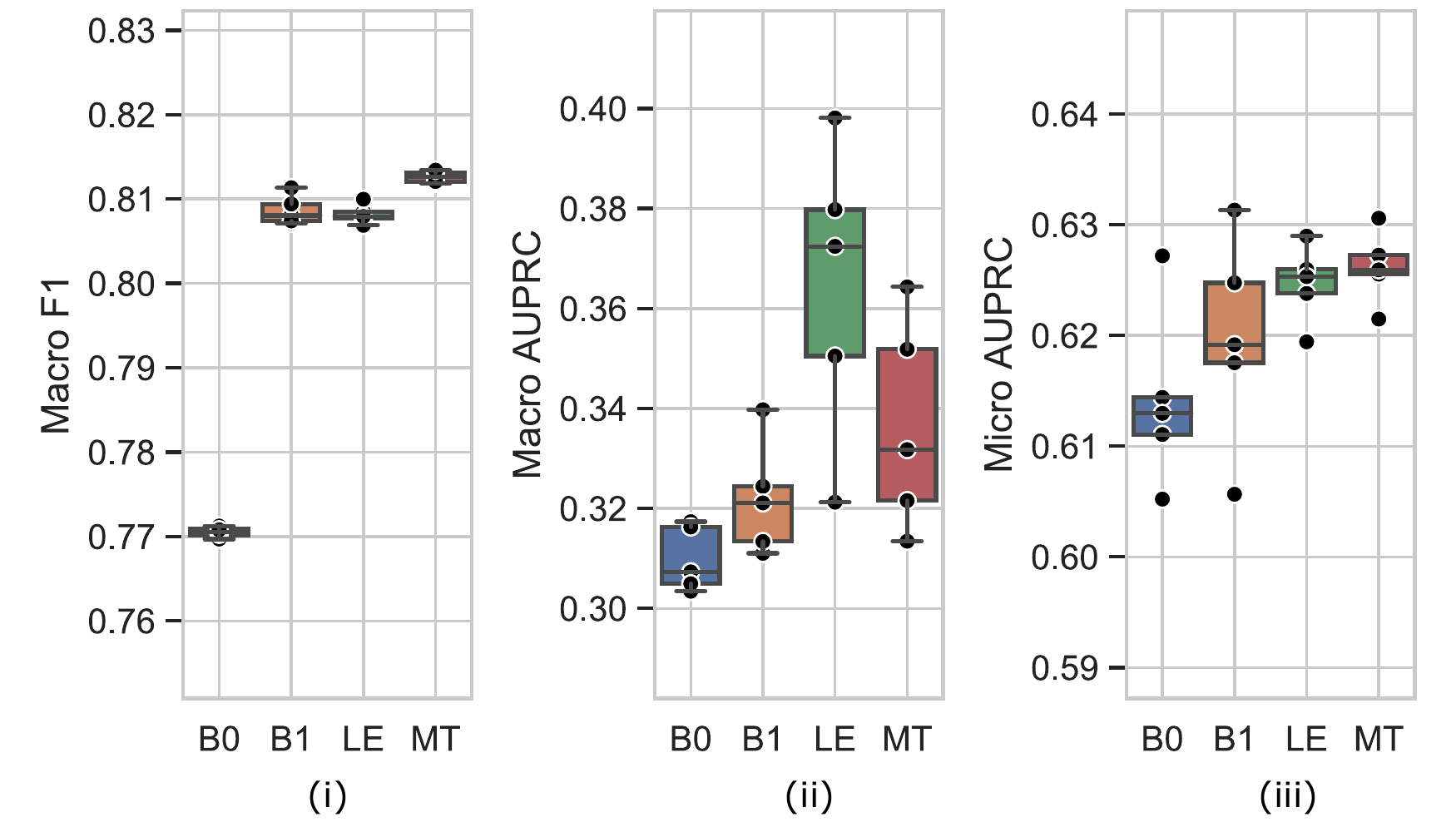}}
    \caption{(i) OpenMIC results, (ii) \& (iii) SONYC-UST results. Each experiment is replicated 5 times to obtain the boxplots.}
    \label{fig:boxplots}
\end{figure}

\Figref{fig:boxplots} confirms that treating all missing labels as explicit negatives~---the B0 approach commonly used in audio classification research---~leads to significantly worse performance compared to the other approaches for both datasets and for almost all runs. B1, on the other hand, is a more competitive baseline. This can be attributed to the fact that it addresses missing labels by omission.

The key difference between the two datasets is that OpenMIC has a small proportion of labeled data while the SONYC-UST has a much larger portion of the dataset labeled. 
While MT outperforms other methods on OpenMIC, there is no clear winner between MT and LE on SONYC-UST. While the average performance of both MT and LE is higher, there are runs where B1 and B0 are able to match their performance. It is likely that LE and MT methods are unable to benefit from the small number of missing labels in SONYC-UST. 
It is worth noting that the hyperparameter $\gamma$ requires tuning for the LE method to perform well. MT, on the other hand, achieves strong performance with suggested hyperparameter settings.

\begin{figure}[!t]
  \begin{minipage}[b]{1.0\linewidth}
    \centering
    \centerline{\includegraphics[width=8cm]{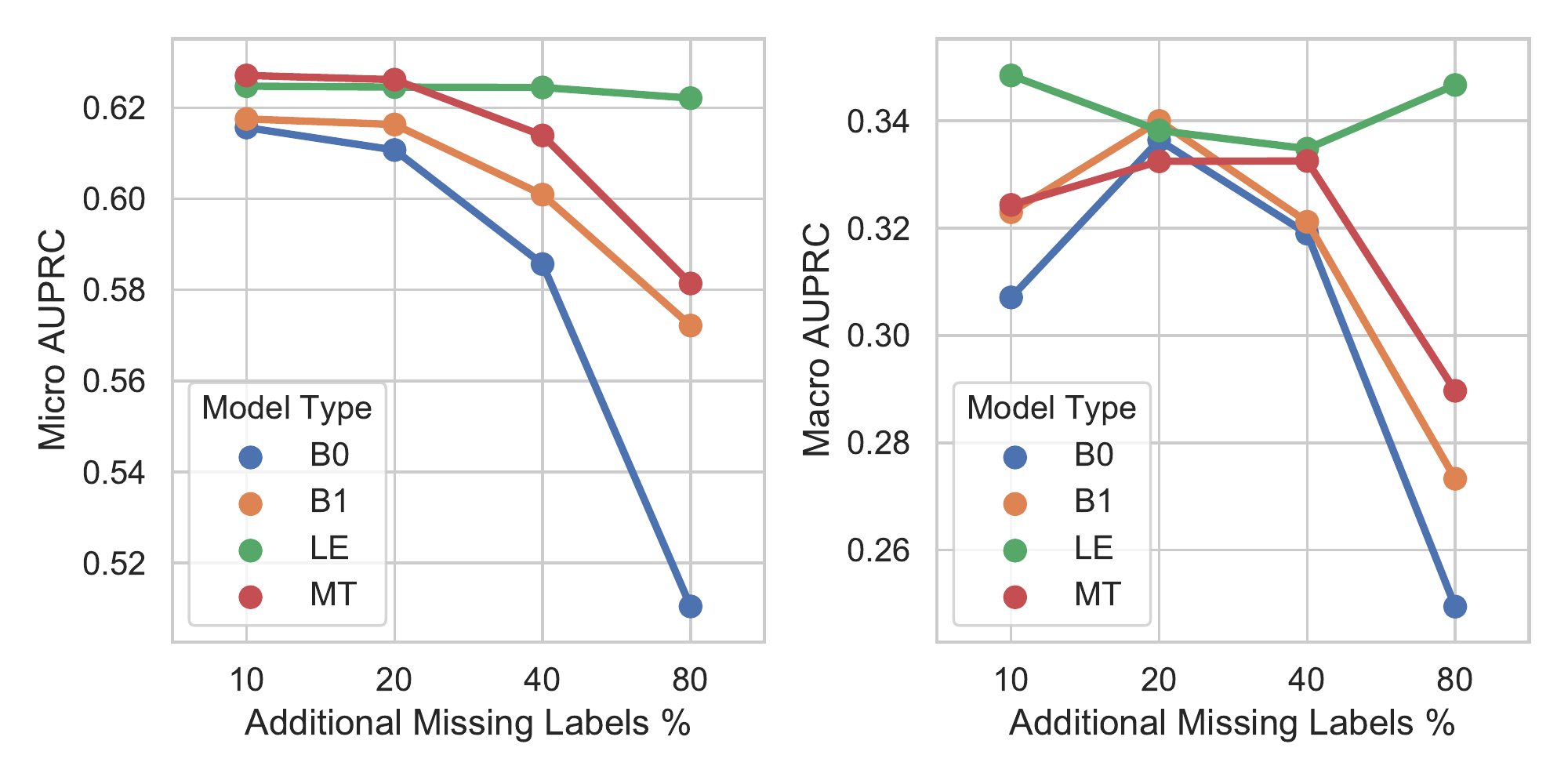}}
    \caption{Effect of increasing the amount of missing labels}
    \label{fig:trend}
  \end{minipage}
\end{figure}

To further investigate, we simulate more missing labels in SONYC-UST. We randomly remove $\{10, 20, 40, 80\}\%$ labels and redo the experiments. Figure~\ref{fig:trend} shows the effect of varying amounts of missing labels on the different algorithms. While it is expected that performance drops with the reduction in labeled information, LE's performance does not drop significantly. The LE method relies on creating an enhanced label set by removing the most confused missing labels, which leads to the student model being trained with potentially more accurate labels. Similarly, MT and B1 do not deteriorate as drastically as B0, reaffirming that even a naive approach of handling missing labels is better than simply treating them as missing.

Note that MT requires that we know which labels are missing. This is a drawback since certain datasets, such as Audioset and MSD, may only have positives and implicit negatives. The LE method is effective in these cases since it attempts to predict which of the implicit negatives are explicit negatives. 

\section{Conclusion and Future Work}

Our paper brings to the fore a common but often overlooked challenge faced in the task of audio classification: partially labeled large-scale data. We show that treating missing labels as negatives, or ignoring/masking missing labels during training often leads to bad performance. LE and MT are both strong candidates for learning with partially labeled data. The main difference between the two is that MT needs to know which labels are missing, while LE does not require this information. Our results demonstrate the ability to leverage missing labels, which has implications on the collection of annotations for audio data. Dataset creation can be simplified since comprehensive annotations are not required and partially labeled data may be sufficient to learn good models for audio classification.
In future work, we plan to study the relationship between the proportion of missing labels and LE's hyperparameter $\gamma$. Additionally, we plan to combine the two approaches, utilizing consistency regularization in both stages of the LE method, potentially leading to cleaner enhanced label sets for the student model.

\bibliographystyle{IEEEtran}
\bibliography{IEEEabrv,ICASSP2021}

\end{document}

%% file: math.tex
\usepackage{amsmath, amsfonts}

\def\Figref#1{Figure~\ref{#1}}

\def\Tabref#1{Table~\ref{#1}}

\def\eqref#1{equation~\ref{#1}}

\def\rvx{{\mathbf{x}}}

\DeclareMathAlphabet{\mathsfit}{\encodingdefault}{\sfdefault}{m}{sl}
\SetMathAlphabet{\mathsfit}{bold}{\encodingdefault}{\sfdefault}{bx}{n}

\newcommand{\expect}[2]{\mathbb{E}_{#1}\left[#2\right]}

